\documentclass[preprint,12pt]{elsarticle}
\usepackage[table]{xcolor}
\usepackage{lscape}
\usepackage{amsmath}
\usepackage{amssymb}
\usepackage{placeins}
\usepackage{ragged2e}
\usepackage{tabulary}
\usepackage{tabularx}
\usepackage{graphicx}
\usepackage{caption}
\usepackage{subcaption}
\usepackage{hhline}
\usepackage{longtable}
\usepackage{rotating}
\usepackage{array}
\usepackage{multicol}
\usepackage{multirow}
\usepackage{booktabs}
\usepackage{cleveref}

\usepackage{tikz}
\def\checkmark{\tikz\fill[scale=0.4](0,.35) -- (.25,0) -- (1,.7) -- (.25,.15) -- cycle;} 

\journal{arXiv}
\begin{document}
\begin{frontmatter}

\title{Factors associated with injurious from falls in people with early stage Parkinson's disease}
\author{Sarini Abdullah$^1$, James McGree$^2$, Nicole White$^{3}$, Kerrie Mengersen$^2$, Graham Kerr$^3$}
\address{$^1$Department of Mathematics, University of Indonesia, Indonesia\\ $^2$ARC Centre of Excellence for Mathematical and Statistical Frontiers, Queensland University of Technology, Australia\\
$^3$Institute of Health and Biomedical Innovation (IHBI), Queensland University of Technology, Australia}
\begin{abstract}
Falls are common in people with Parkinson's disease (PD) and have detrimental effects which can lower the quality of life.  While studies have been conducted to learn about falling in general, factors distinguishing injurious from non-injurious falls are less clear. We develop a two-stage Bayesian logistic regression model was used to model the association of falls and injurious falls with data measured on patients. The forward stepwise selection procedure was used to determine which patient measures were associated with falls and injurious falls, and Bayesian model averaging (BMA) was used to account for uncertainty in this variable selection procedure.  

Data on 99 patients for a 12-month time period were considered in this analysis. Fifty five percent of the patients experienced at least one fall, with a total of 335 falls cases; 25\% of which were injurious falls.
Fearful, Tinetti gait, and previous falls were the risk factors for fall/non-fall, with 77\% accuracy, 76\% sensitivity, and 76\% specificity. Fall time, body mass index, anxiety, balance, gait, and gender were the risk factors associated with injurious falls. Thus, attaining normal body mass index, improving balance and gait could be seen as preventive efforts for injurious falls. There was no significant difference in the risk of falls between males and females, yet if falls occurred, females were more likely to get injured than males. 

\end{abstract}
\end{frontmatter}
\section{Introduction}

Falls are common in people with Parkinson’s disease (PD), and have great clinical significance that adversely affecting the patient's health \cite{Roller1989}.  
Although some studies have investigated which factors are associated with falls, there is still little known about the factors distinguishing injurious from non-injurious falls in PD. Talking while walking \cite{LaPointe2010}, being female and older age \cite{Wielinski2005} were found to be positively associated with injurious falls. 

Despite these findings, there is still the need for a comprehensive study for prediction of falls and injury from falling in people with PD. Identification of injurious falls risk factors based only on one fall occurrence on each patient may not sufficiently provide enough information on factors driving injurious falls. Ideally, information on factors that are possibly associated with injurious falls would be better preserved in a repeated observation of falls, as will be showcased in this paper. Identification of these factors could have great clinical relevance for PD treatment. If predictive characteristics of the faller or the fall associated with injury could be identified, preventive measures could be targeted toward these characteristics and individuals. 

Logistic regression model is commonly applied to predict falls in PD \cite{Allcock2009, Kerr2010, Pickering2007, Sun2009, Wood2002}. However, several conditions are required in order to establish the usefulness of a logistic regression model. One guideline suggests that there should be at least 10 participants for each predictor \cite{Agresti2007} which is often difficult to fulfill in many PD data applications, due to the limited data size. Moreover, the determination of the best subset of predictors for inclusion is challenging. Finally, predictions based on the set of selected variables do not take into account the uncertainty associated with this selection. 

A common selection criteria for variable selection is based on the likelihood ratio test (LRT) \cite{Neyman1928, Neyman1928a}. A (set of) variable is said to contribute to the model if the inclusion of that (set of) variable changes the likelihood ratio between the resulting model and the current model. Thus, it is often assumed that the selection process is done in the framework of nested models, i.e. one model is a special case of another. However, there are many situations in which it is important to compare non-nested models, and so, in these cases , this would be a substantial drawback of using LRTs for model comparison. Nevertheless, in reality it is actually possible to use the LRT for comparing both nested and non-nested models \cite{Cox1962}. Backward and forward selection are the commonly used procedure for variable selection based on the LRT. However, there are issues of multiple testing when performing a series of LRTs which can result in inflated false discovery rates \cite{Benjamini1995}. 

In this paper we address the shortcomings by adopting a Bayesian approach to variable selection. The log marginal likelihood is, a well established model selection criterion in Bayesian statistics \cite{Hubin2016}, and could be seen as an alternative statistic for variable selection. Such uncertainty can be accounted for by a method called Bayesian model averaging (BMA) \cite{Hoeting1999}, where (weighted) predictions from different models are averaged across all potential models (or subsets of the predictors). This method is useful in this study as uncertainty in the association between factors and falls, and injurious falls will still remain, even though data have been collected on 99 PD patients. 

The aim of this paper is to identify fall risk factors in people with PD, and given they fell, identify risk factors associated with injurious/non-injurious falls. This will be achieved via a two-stage Bayesian logistic regression model. To account for repeated measures on each patient, corresponding to multiple occurrences of falls, a regression model with random effects is proposed. To the best of our knowledge, this is the first study in Parkinson's that fits simultaneously modelling falls/non-falls and injured/non-injured. The rest of the paper is organized as follows. A description on data and statistical methods is given in Section \ref{methods}. Section \ref{Sec:result} presents the key findings of this study, followed by a discussion in Section \ref{sec:discussion}. Finally, a summary of overall findings is given in Section \ref{sec:summary}. 

\section{Data and Methods}
\label{methods}
\subsection{Data} \label{sec:data}
\justify

\textbf{Participants.} 99 people diagnosed with idiopathic PD were prospectively recruited from community support groups and neurology clinics in southeast Queensland from March 2002 until December 2006. This recruitment was a part of a larger research project conducted by the Institute of Health and Biomedical Innovation in Brisbane, Australia \cite {Kerr2010}. All participants were classified as early stage PD, determined by a Hoehn and Yahr (HY) score of 3 or less. 

\textbf{Assessments.} Each participant completed a monthly falls diary over a consecutive twelve month period. Prospectively, participants were classified as fallers if they recorded any falls during follow-up. Successful completion of the diary was monitored by phone calls and mail correspondence. 

A series of clinical and functional tests were conducted at baseline.Functional tests were Tinetti, timed up and go (TUG), and functional reach (FR). The Tinetti test comprised 2 subscales, which relate to clinical balance and gait. Patients' quality of life was measured by the 36-item Short Form Health Survey (SF-36) which represented 8 subscales: physical functioning, role limitations due to physical health, role limitations due to emotional problems, energy/fatigue, emotional well-being, social functioning, pain, and general health. A self report rating on balance, fearfulness, falls experience in the past year, depression (using Beck's Depression Inventory, BDI) and anxiety (using the Beck's Anxiety Inventory, BAI) were also recorded. For each fall occurrence, the time and location of falls, and whether patients were wearing glasses was also recorded.

\subsection{Statistical methods}
\subsubsection{Logistic regression with random effects} 

In logistic regression, a single outcome variable $Y_i, i = 1,..., n$ follows a Bernoulli
distribution that takes on the value 1 with probability $\pi_i$ and 0 with probability
$1-\pi_i$. The log odds, $\frac{\pi_i}{1-\pi_i}$ is then modelled as a linear function of explanatory variables. In this Section, a two-stage logistic regression is developed to predict falls (Stage 1) and injury associated with falls (Stage 2). 

Let $Y_{1i} = 1$ if patient $i$ fell, and $0$ otherwise. Using $P$ predictor variables and including $1$ for the intercept term, $\textbf{X}_{1i}=(1, x_{1i}, ..., x_{Pi})$, model for Stage 1 can be written as \citep{McCullagh1989}:
\begin{equation}\label{eq:m1}
Y_{1i} \sim \text{Bernoulli} \,(\pi_{1i}),
\end{equation}
\begin{equation}
\text{logit} (\pi_{1i})=\beta_{0}+\beta_{1}x_{1i}+...+\beta_{P}x_{Pi}
\end{equation}  
with $\boldsymbol{\beta}=(\beta_0, ..., \beta_P)$ are the regression coefficients. For the $p$-th covariate, its contribution in the model is quantified through the odds ratio of falling, $\text{exp}({\beta_p}), p=1,...,P$.

If patient $i$ had multiple falls, say $R$ falls, let $Y_{2ir} = 1$ if patient $i$ had an injurious fall at the $r$-th fall and $0$ otherwise, $r=1,...,R$. In addition to the $P$ variables potentially associated with fall/non-fall, we consider $L$ variables which are potentially associated with whether the patient was injured or not. Let $\textbf{X}_{2i}$ denote the predictor variables for patient $i$ for the Stage 2 model, that is, $\textbf{X}_{2i}=(\textbf{x}_{(K+1)i}, ..., \textbf{x}_{(K+L)i})$, with $\textbf{x}_{(K+l)i}=(x_{(K+l)i1}, ..., x_{(K+l)iR})$. Moreover, to account for the specific effects of patient $i$, the random effects term $\epsilon_i$ is also added to the model. 
Thus,  
\begin{equation}\label{eq:m2}
Y_{2ir} \sim \text{Bernoulli} \,(\pi_{2ir}),
\end{equation}
\begin{equation}\label{eq:model2}
\text{logit} (\pi_{2ir})= \textbf{x}_{1i} \boldsymbol{\alpha}_1^T+\textbf{x}_{2i} \boldsymbol{\alpha}_2^{T}+\epsilon_i, \quad \epsilon_i \sim N(0,\sigma^2)
\end{equation}  
with $\textbf{x}_{i} \boldsymbol{\alpha}_1^T = \alpha_0 + \alpha_1 x_{1i}+...+\alpha_K x_{Ki}$ is the contribution associated with the patient measures at baseline and $\textbf{x}_{2i} \boldsymbol{\alpha}_2^{T}=\alpha_{K+1} x_{(K+1)ir}+...+\alpha_{K+L} x_{(K+L)ir}$ is the contribution associated with measurements at each fall.
 
Within a Bayesian framework, one needs to specify the prior distribution for the model parameters. Here, each regression coefficient is assumed to follow a Gaussian distribution, $\beta_k \sim N(0,v_0),$ and $\alpha_l \sim N(0,v_0)$ for $k=0,...,K, l=0,...,L$.  While for the random effects variance, $\sigma^2 \sim IG(a,b)$ where $IG(a,b)$ is an Inverse Gamma distribution with shape $a>0$ and rate $b>0$. We set the regression coefficients variance $v_0=1,000$, and $a=b=0.001$ to represent vague prior knowledge.

Let $\textbf{Z}=\{\textbf{X}_1,\textbf{X}_2\}$ and $\boldsymbol{\theta}=\{\boldsymbol{\alpha}_1, \boldsymbol{\alpha}_2\}$. The posterior distribution of the parameters is then defined as
\begin{equation}
p(\boldsymbol{\beta},\boldsymbol{\theta},\sigma^2|\textbf{Y}_1, \textbf{Y}_2, \textbf{Z}) \propto p(\textbf{Y}_1|\textbf{X}_1, \boldsymbol{\beta})p(\textbf{Y}_2|\textbf{Z},\boldsymbol{\theta},\sigma^2)p(\boldsymbol{\beta})p(\boldsymbol{\theta})p(\sigma^2), 
\end{equation}  
where $p(\textbf{Y}_1|\textbf{X}, \boldsymbol{\beta})$ and $p(\textbf{Y}_2|\textbf{Z},\boldsymbol{\theta},\sigma^2)$ are the likelihood of data $\textbf{Y}_1$ and $\textbf{Y}_2$ as given in Equations (\ref{eq:m1}) and (\ref{eq:m2}) for Stage 1 and Stage 2 of the model, respectively. $p(\boldsymbol{\beta})$, $p(\boldsymbol{\theta})$, and $p({\sigma^2})$ are the prior distributions of $\boldsymbol{\beta}, \boldsymbol{\theta}$, and ${\sigma^2}$, respectively. The posterior distribution was approximated using the Integrated nested Laplace approximation (INLA) \citep{Rue2009}, and is implemented using the open-source software R-INLA package that can be found on www.r-inla.org.

A forward variable selection procedure using the log marginal likelihood (\textit{lml}) as a criterion. The likelihood measures how the model fits the data \cite{Gelman2014}, and marginalizing it over the set of parameters produces a marginal likelihood. By the marginalizing process, it accounts for all the model's parameters which implies a trivial inbuilt penalty for model complexity. It is difficult to calculate the marginal likelihood analytically as most of the models contain unknown parameters, and thus an approximation is required. Among many approaches to approximate the marginal likelihood, Integrated nested Laplace approximation (INLA) \cite{Rue2009} has become a popular choice for its computationally fast yet still reasonably precise \cite{Hubin2016}. 

In a forward variable selection procedure, variables were added to the model one at a time. Starting with a model consisting of an intercept term only, at each step, each variable that was not in the model was tested for inclusion in the model. The variable that yielded the highest $lml$ in that stage was included in the model, as long as the $lml$ was higher than that of the current model. The process continued until there was no more increase in the $lml$. In this paper, we conducted the variable selection for Models 1 and 2 separately. Thus, it allows for the variables that were not selected in the model in Stage 1 to be considered again in Stage 2. 
 
While the forward variable selection procedure should yield a selection of important variables, it ignores model uncertainty. To overcome this problem we considered several potential models, where the prediction was made by averaging the results from these models. All models that were considered during the forward procedure were fitted, and predictions were made for each model. A final prediction, called BMA prediction, was then calculated by the weighted average of predictions from all considered models, with the ratio of the model's $lml$ to the total $lml$ of all models as the weights. Data processing for this paper was conducted in R version 3.2.5 \cite{RCoreTeam2016}.

\subsection{Model assessment}

Bayesian model averaged predictions were assessed in terms of their ability to predict new data via the leave-one-out cross validation method. For each observation $i$, its predicted value is calculated via a model that was fit with all data except the $i^{th}$ observation. 
The classification rates were in the form of sensitivity (or true positive rate, TPR), specificity (1-false positive rate, FPR), and accuracy. A sensitivity is the proportion of correctly classifying true fallers in stage 1 (or injurious falls in stage 2), a specificity is the proportion of correctly classifying true non-fallers in stage 1 (or non-injurious falls in stage 2), and an accuracy is the proportion of correctly classifying true fallers and non-fallers in stage 2 (or true injurious and non-injurious falls in stage 2). 

While it is trivial to calculate the predicted probability once the regression coefficients, $\beta_j$s, were estimated in Stage 1, it is more complex in Stage 2 since it includes the random effects (see Equation \ref{eq:model2}). To predict whether a patient will have an injurious fall, a sample of $\sigma^{2}$ is drawn from its posterior distribution. For each of $\sigma^{2}$, the random effect is drawn from $N(0,\sigma^2)$, and then added to the fixed component of the model to obtain the predicted log-odds of getting injurious falls. Averaging these values over sample and exponentiating it, we obtain the predicted probability of injurious falls for the patient. 

Once the predicted probabilities are obtained, the classification is based on a chosen threshold (ranging from 0 to 1) that jointly optimises the sensitivity and specificity. ROC curves and the corresponding area under the ROC curves (AUC) were also presented. A useful model will have an ROC curve that is close to the upper left corner of the graph, and thus the corresponding AUC will close to 1. Graphs of ROC curves in this paper were produced using ROCR package \cite{Sing2005}.

\section{Results}
\label{Sec:result}
\subsection{Participants description and data exploratory}

A total of 99 patients were included in the analysis, with 55 of them experiencing at least one fall during the 12-month period. Among these fallers, 20 fell once, and the remaining were recurrent fallers, giving a total of 335 falls. Around 25\% of falls cases ended with injuries, with various types of injuries as listed in Table \ref{tab:injury}. Most falls caused bruises, varying from mild to severe cases. Other examples of types of injuries were abrasion (face, arm, leg), swelling, and sprained and broken wrist.

\begin{table}[htbp]
  \centering
  \caption{Various injuries resulting from falls in PD patients.}
    \scalebox{0.8}{
		\begin{tabular}{lc}
    \toprule
    \rowcolor[rgb]{ .949,  .949,  .949} \textbf{Types of injuries} & \textbf{Frequency} \\
    \midrule
    \midrule
    Bruise & 40 \\
    Injuries (toes, knees, hands, arms) & 21 \\
    Abrasion \& broken skin & 9 \\
    Hit head & 6 \\
    Cuts (arm, finger, knee) & 5 \\
    Jarred (hands, knees, wrist) & 5 \\
    Laceration & 5 \\
    Swelling & 4 \\
    Sprained (wrist, vertebrae) & 4 \\
    Broken wrist & 3 \\
    Facial bruising & 2 \\
    Twisted (ankle,  knee) & 2 \\
    Pulled muscles & 1 \\
    \bottomrule
    \end{tabular}%
		}
  \label{tab:injury}%
\end{table}%

A summary of the variables used in the analysis is given in Table \ref{tab:descriptive}. Patients were classified as fallers/non-fallers, and if they fell, they were further classified as injured or non-injured. A univariate between group test was conducted using the Mann-Whitney-Wilcoxon test for each continuous variable, and a chi-square test for categorical variables. A p-value $< 0.05$ implies such variable significantly differentiates the two groups. 

\begin{table}[htbp]
  \centering
  \caption{Summary of variables for all patients (mean for continuous variable, count for categorical variable), and classified by fall/ not-fall (for Stage 1), and injured/not-injured (for Stage 2). The p-value is the statistical significance for the Mann-Whitney-Wilcoxon test (for continuous variables) and the chi-square test (for categorical variables). In Stage 1, the count is for the patient number, and it is the fall number (\%) in Stage 2. The percentages add up to 100\% over categories for All patients, for fall/not-fall (or injured/not-injured) within each category for Stage 1 (Stage 2).} 
	\scalebox{0.7}{
    \begin{tabular}{l|c|ccc|ccc}
    \multicolumn{1}{r}{} & \multicolumn{1}{r}{} &      &      & \multicolumn{1}{r}{} &      &      &  \\
    \midrule
    \rowcolor[rgb]{ .949,  .949,  .949} \multicolumn{1}{c|}{\textbf{Measurement}} & \textbf{All } & \multicolumn{3}{c|}{\textbf{Stage 1}} & \multicolumn{3}{c}{\textbf{Stage 2}} \\
\cmidrule{3-8}    \rowcolor[rgb]{ .949,  .949,  .949}      & \textbf{$(N=99)$} & \textbf{Fall} & \textbf{Not-fall} & \textbf{p-value} & \textbf{Injured} & \textbf{Not-injured} & \textbf{p-value} \\
    \midrule
    \midrule
    Age  & 66.9 & 67.3 & 66.5 & 0.62 & 66.6 & 67.3 & 0.34 \\
    Tinetti &      &      &      &      &      &      &  \\
\quad       Balance & 14.9 & 14.6 & 15.3 & 0.02 & 13.9 & 14.2 & 0.32 \\
\quad        Gait & 10.6 & 10.3 & 10.9 & 0.01 & 9.5  & 10.5 & $<.001$ \\
    Functional Reach  & 28.3 & 28.6 & 27.8 & 0.55 & 27   & 28.2 & 0.12 \\
    Timed Up and Go  & 111.2 & 10.7 & 236.8 & 0.32 & 11.4 & 12   & 0.06 \\
    Previous falls & 0.8  & 1.1  & 0.5  & 0.02 & 1.8  & 1.05 & 0.001 \\
    SF-36 &      &      &      &      &      &      &  \\
     \quad  Physical functioning & 73.4 & 68.3 & 79.9 & 0.01 & 58.8 & 37.8 & $<.001$ \\
      \quad Physical health & 58.8 & 49.5 & 70.5 & 0.01 & 42.1 & 19   & $<.001$ \\
      \quad Bodily pain & 74.5 & 73.5 & 75.8 & 0.59 & 68.5 & 70.3 & 0.52 \\
      \quad General health & 61.2 & 60.2 & 62.4 & 0.58 & 52.9 & 37.1 & $<.001$ \\
      \quad Vitality & 57.8 & 54.7 & 61.6 & 0.08 & 47.7 & 32.2 & $<.001$ \\
      \quad Social functioning & 83.5 & 82.7 & 84.4 & 0.7  & 73.3 & 59.7 & $<.001$ \\
      \quad Emotional problems & 81.5 & 77   & 87.1 & 0.12 & 59.3 & 30.2 & $<.001$0 \\
      \quad Mental health & 81.1 & 79.1 & 83.5 & 0.14 & 72.9 & 53.7 & $<.001$ \\
    Beck's Depression  & 6.1  & 7    & 5    & 0.05 & 8.2  & 9.4  & 0.15 \\
    Beck's Anxiety  & 6.8  & 8.2  & 5.2  & 0.01 & 8.3  & 12   & $<.001$ \\
    \midrule
    \midrule
    Gender   &      &      &      & 0.67 &      &      & 0.01 \\
    \quad     Male & 43 (43.4\%) & 25 (58.1\%) & 18 (41.9\%) &      & 35 (16.1\%) & 182 (83.9\%) &  \\
        \quad Female & 56 (56.6\%) & 30 (56.3\%) & 26 (46.4\%) &      & 43 (38.4\%) & 69 (61.6\%) &  \\
    Body mass index  &      &      &      & 0.35 &      &      & 0.01 \\
    \quad Normal & 44 (44.4\%) & 20 (45.5\%) & 24 (54.5\%) &      & 21 (11.1\%) & 168 (88.9\%) &  \\
    \quad Overweight & 34 (34.3\%) & 20 (58.8\%) & 14 (41.2\%) &      & 36 (39.1\%) & 56 (60.9\%) &  \\
    \quad Obese & 21 (21.2\%) & 15 (71.4\%) & 6 (28.6\%) &      & 21 (43.8\%) & 27 (56.3\%) &  \\
    Previous falls   &      &      &      & 0.05 &      &      & $<.001$ \\
    \quad Yes  & 37 (37.4\%) & 27 (73\%) & 10 (27\%) &      & 53 (39.6\%) & 81 (60.4\%) &  \\
    \quad No   & 62 (62.6\%) & 28 (45.2\%) & 34 (54.8\%) &      & 25 (13\%) & 168 (87\%) &  \\
    Balance &      &      &      & 0.04 &      &      & $<.001$ \\
    \quad Excellent & 4 (4\%) & 2 (50\%) & 2 (50\%) &      & 1 (100\%) & 0    &  \\
    \quad Very good & 28 (28.3\%) & 12 (42.9\%) & 16 (57.1\%) &      & 13 (38.2\%) & 21 (61.8\%) &  \\
    \quad Good & 35 (35.4\%) & 19 (54.3\%) & 16 (45.7\%) &      & 23 (38.3\%) & 37 (61.7\%) &  \\
    \quad Fair & 27 (27.3\%) & 18 (66.7\%) & 9 (33.3\%) &      & 36 (16.1\%) & 188 (83.9\%) &  \\
    \quad Poor & 5 (5.1\%) & 4 (80\%) & 1 (20\%) &      & 5 (50\%) & 5 (50\%) &  \\
    Fearful &      &      &      & 0.21 &      &      & $<.001$ \\
    \quad Not at all & 40 (40.4\%) & 18 (45\%) & 22 (55\%) &      & 16 (36.4\%) & 28 (63.6\%) &  \\
    \quad Slightly & 40 (40.4\%) & 21 (52.5\%) & 19 (47.5\%) &      & 25 (39.7\%) & 38 (60.3\%) &  \\
    \quad Moderately & 11 (11.1\%) & 8 (72.7\%) & 3 (27.3\%) &      & 6 (18.8\%) & 26 (81.3\%) &  \\
    \quad Quite a bit & 6 (6.1\%) & 6 (100\%) & 0    &      & 29 (15.8\%) & 155 (84.2\%) &  \\
    \quad Extremely & 2 (2\%) & 2 (100\%) & 0    &      & 2 (33.3\%) & 4 (66.7\%) &  \\
    \bottomrule
    \end{tabular}%
}
  \label{tab:descriptive}%
\end{table}%

Several variables that did not significantly differentiate fallers and non-fallers were able to differentiate injured from non-injured patients, and vice versa. Tinetti (balance and gait), previous falls, physical functioning and physical health (both as measured by the SF-36), depression and anxiety (BDI, BAI) were independently associated with fall/non-fall. However, balance and depression were not associated with injurious falls. In contrast, general health, social functioning, emotional problems, and mental health (all as measured by the SF-36) differentiated those with injurious falls from those with non-injurious falls despite their non-significance in differentiating fallers from non-fallers. The proportion of fall cases is almost the same for males and females. Falls seem more likely in patients with high BMI (overweight or obese), and those who had experienced falls prior to the study. It is worth noting that patients with an excellent self-rating assessment on balance appear to have the same likelihood of falls and non-falls. 

The Stage 2 columns of Table \ref{tab:descriptive} present the proportion of cases with injurious/non-injurious falls. A relatively higher proportion of injurious falls is observed for females than males, for patients who were overweight and obese, and patients who had at least one fall prior to the study time. Patients with high confidence in their balance (good, very good, and excellent) appeared more likely to have injurious falls than those with fair balance. Similarly for the self-rated assessment on fearful to fall, a higher proportion of injurious falls was obtained for those who were not at all or slight fearful of falls than in those with moderate or quite a bit fear of falls. 
\begin{figure}[t!]
\centering
\begin{subfigure}[b]{.56\textwidth}
  \centering
  \includegraphics[keepaspectratio=true,scale=0.35]{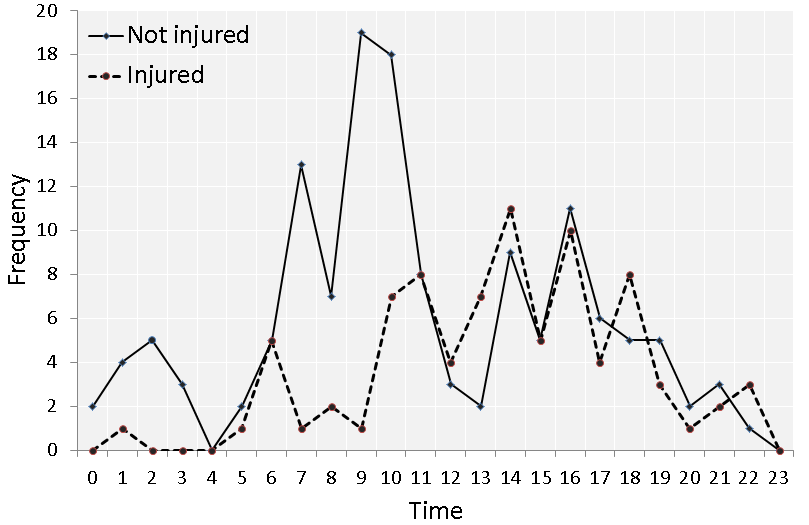} 
  \caption{}
  \label{fig:timeA}
\end{subfigure}%
~
\begin{subfigure}[b]{.4\textwidth}
  \centering
  \includegraphics[keepaspectratio=true,scale=0.48]{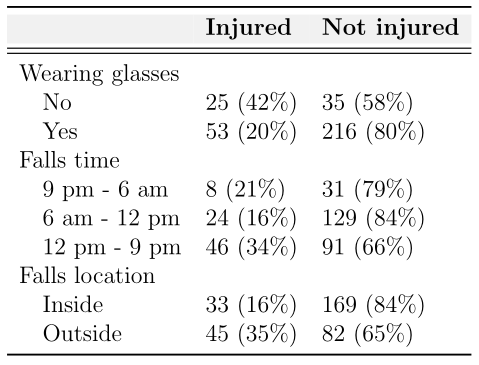} 
  \caption{}
  \label{fig:timeB}
\end{subfigure}
\caption{Distribution of fall times (24-hour time) (a), and summary of variables specific in each fall occurrence classified by injured/not-injured (b).}
\label{fig:time}
\end{figure}

The distribution of fall times is presented in Figure \ref{fig:timeA}. Falls mostly occurred during the day, particularly in the morning between 6 a.m. - 11 a.m. Injurious falls were apparently more likely to occur in the afternoon. Injurious falls reached a peak rate at 1 p.m., where around 80\% of falls caused injury. It is also worth noting that the frequency of injurious falls was higher than that of non-injurious falls at night time, particularly at 6 p.m. and 10 p.m. 

Based on this time distribution, we partitioned falls time into three time intervals (Figure \ref{fig:timeB}). It is shown that although falls mostly occurred in the morning, only 16\% of these cases caused injury. For falls that occurred between 12 p.m. and 9 p.m., the rate of injurious falls rose to 34\%. With regard to the location, injurious falls seem to be more likely to occur outside than inside the house.


\subsection{Stage 1 results: Falls risk factors}

\begin{figure}[h]
\centering\includegraphics[width=0.5\linewidth]{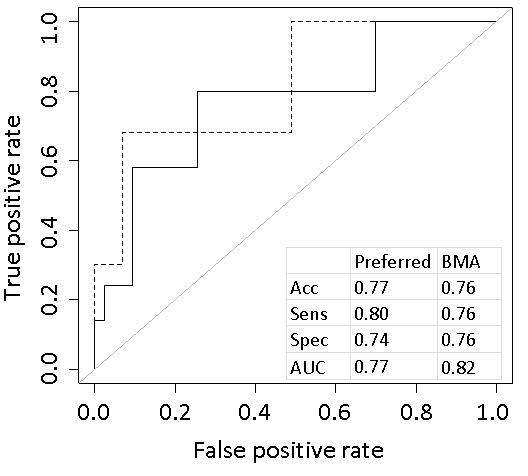}
\caption{ROC curves for fall/non-fall classification based on the best model (solid) and BMA models (dashed).}
\label{fig:roc1}
\end{figure}

Our adopted forward variable selection procedure yielded three variables as important predictors for falls: fearfulness, Tinetti gait, and previous falls. Fear of falling was positively associated with the risk of experiencing a fall. Patients experiencing falls prior to participation in the study had a 2.5-fold higher risk of falling than the new fallers. Improvement in gait was associated with a lower risk of fall, at 0.59, with the 95\% CI (between 0.39 and 0.85 times less risk) for each 1 point reduction in the Tinetti gait score. The selected model correctly identified fallers with the sensitivity of 80\%, while for non-fallers, it was a bit lower at the specificity of 74\% (Figure \ref{fig:roc1}).  

Classification rates and ROC analysis for BMA are presented in Figure \ref{fig:roc1}. The difference with the preferred model is small in terms of the accuracy, sensitivity, and specificity. The AUC for BMA is slightly higher than that from the preferred model, implying an improvement using BMA, that is, taking into account model uncertainty. Five models with relatively high posterior model probabilities considered in BMA are listed in Table \ref{tab:BMA1}. Other factors that also have the potential to explain falls are anxiety (measured by BAI) and the physical functioning sub-scale of the SF-36. 

\begin{table}[htbp]
  \centering
  \caption{Variables that were selected in five models with the highest weight to predict fall/non-fall based on BMA procedure.}
    \scalebox{0.8}{
				\begin{tabular}{lccccc}
    \toprule
    \rowcolor[rgb]{ .949,  .949,  .949} \textbf{Variable} & \multicolumn{5}{c}{\textbf{Model}} \\
\cmidrule{2-6}    \rowcolor[rgb]{ .949,  .949,  .949}      & \textbf{1} & \textbf{2} & \textbf{3} & \textbf{4} & \textbf{5} \\
    \midrule
    \midrule
    Fearful & \multicolumn{1}{l}{\checkmark} & \multicolumn{1}{l}{\checkmark} &      &      &  \\
    Tinetti gait & \multicolumn{1}{l}{\checkmark} &      & \multicolumn{1}{l}{\checkmark} & \multicolumn{1}{l}{\checkmark} &  \\
    Previous falls & \multicolumn{1}{l}{\checkmark} & \multicolumn{1}{l}{\checkmark} & \multicolumn{1}{l}{\checkmark} &      &  \\
    BAI  &      &      &      & \multicolumn{1}{l}{\checkmark} &  \\
    Physical functioning &      &      &      &      & \multicolumn{1}{l}{\checkmark} \\
    \midrule
    \rowcolor[rgb]{ .949,  .949,  .949} \textbf{Weight} & 0.45 & 0.21 & 0.14 & 0.05 & 0.04 \\
    \bottomrule
    \end{tabular}%

		}
  \label{tab:BMA1}%
\end{table}%


\subsection{Stage 2 results: Injurious falls risk factors}

Having identified risk factors for falls, further analysis was undertaken to identify factors associated with injury/non-injury. Upon fitting the hierarchical logistic regression model with the random intercept to accommodate the possibility of a patient's specific effects (see Equation (\ref{eq:model2})), the results obtained are as follows. 

\begin{figure}[htbp]
\centering
\begin{subfigure}[b]{.5\textwidth}
  \centering
  \includegraphics[keepaspectratio=true,scale=0.38]{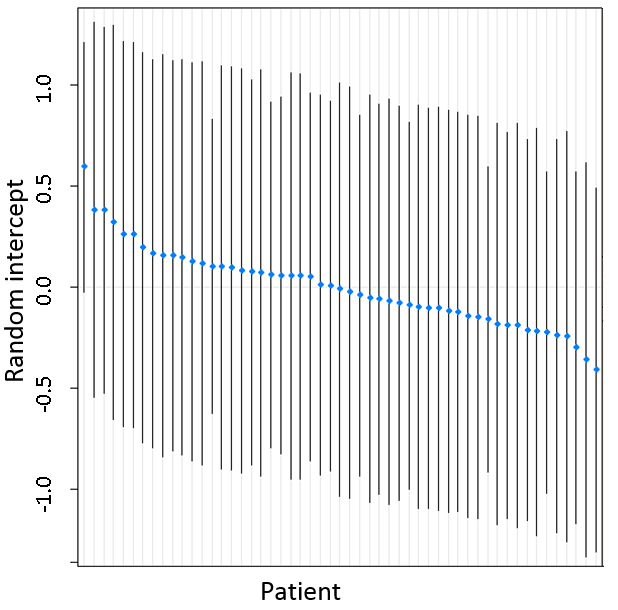} 
  \caption{}
  \label{fig:ch4_randomIntercept}
\end{subfigure}%
~
\begin{subfigure}[b]{.4\textwidth}
  \centering
  \includegraphics[keepaspectratio=true,scale=0.5]{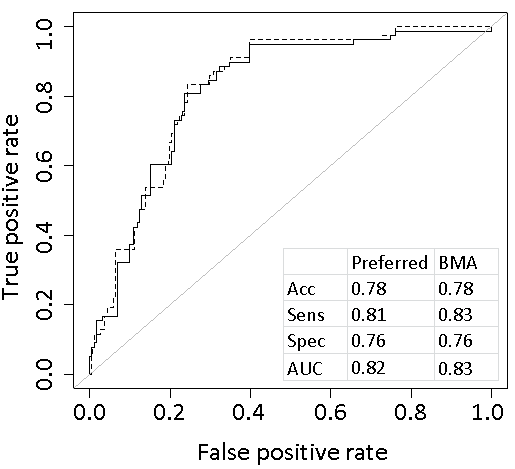} 
  \caption{}
  \label{fig:roc2}
\end{subfigure}
\caption{Random intercepts (with the 95\% credible intervals) based on the preferred model (a), and ROC curves for injurious/non-injurious falls classification based on the preferred model (solid) and BMA models (dashed) (b).}
\label{fig:InterceptAndROC}
\end{figure}

Figure \ref{fig:ch4_randomIntercept} shows that the random intercepts (sum of the intercept and the random effects) for all patients are similar, as 0 is within the $95\%$ credible intervals of the random intercepts. This implies that there is no inherent specific-patient effects in injurious/non-injurious falls. This means that there are no patients having a significantly high or low rate of injurious falls that is not accounted for by the covariates.

Factors that were associated with injurious falls were identified. Four sub-scales of the SF-36: physical functioning, general health, energy/fatigue, and pain were slightly associated with injurious falls although they did not greatly affect the risk of getting an injurious fall. An injury is more likely to occur in patients with a higher balance score (measured by Tinetti balance), shorter functional reach, fearfulness to falls, and for falls that occurred in the afternoon, outside the house.   

Higher Tinetti balance scores were associated with a higher risk of injury from a fall. On average, a one unit increase in Tinetti balance was associated with 1.82 times the risk of getting injured. Fall time seems to be highly associated with injury. The risk of getting injured if falls happened between 6 a.m. - 12 p.m. is 0.2 (CI: 0.09 - 0.40) , significantly lower than that between 12 p.m. - 9 p.m. That said, the risk of injurious falls occuring in the morning (6 a.m. - 12 p.m.) is around 20\% (with the 95\% CI between 9\% and 40\%) of that during 12 p.m. - 9 p.m. For falls that occurred inside a house, the risk of getting injured was half for if they occurred outside. 

\begin{table}[h]
  \centering
\caption{The selected variables for injurious/non-injurious falls classification from five models with relative high weights.}
	\scalebox{0.80}{
    \begin{tabular}{lccccc}
    \toprule
    \rowcolor[rgb]{ .949,  .949,  .949} \textbf{Variable} & \multicolumn{5}{c}{\textbf{Model}} \\
\cmidrule{2-6}    \rowcolor[rgb]{ .949,  .949,  .949}      & \textbf{1} & \textbf{2} & \textbf{3} & \textbf{4} & \textbf{5} \\
    \midrule
    \midrule
    \textbf{SF 36} &      &      &      &      &  \\
     \quad   Physical functioning & \checkmark    & \checkmark    & \checkmark    & \checkmark    & \checkmark \\
     \quad    General health & \checkmark    &      &      &      &  \\
        \quad Energy/fatigue & \checkmark    &      &      &      &  \\
        \quad Pain & \checkmark    &      &      &      &  \\
    \textbf{Tinetti} &      &      &      &      &  \\
        \quad Balance & \checkmark    &      &      &      &  \\
        \quad Gait &      &      & \checkmark    &      &  \\
    FRB  & \checkmark    &      &      & \checkmark    &  \\
    Fall time & \checkmark    & \checkmark    & \checkmark    & \checkmark    & \checkmark \\
    Fall location & \checkmark    &      &      &      &  \\
    Fearful & \checkmark    &      &      &      &  \\
    BMI  &      &      &      &      & \checkmark \\
    BAI  &      & \checkmark    &      &      &  \\
    Gender  &      &      &      & \checkmark    &  \\
    \midrule
    \rowcolor[rgb]{ .949,  .949,  .949} \textbf{Weight} & 0.45 & 0.21 & 0.14 & 0.05 & 0.04 \\
    \bottomrule
    \end{tabular}%
}
  \label{tab:bma}%
\end{table}%

The results from BMA are similar to those from the preferred model in terms of the classification rates. A list of potential risk factors for injurious falls from five models with relatively high posterior model probability is presented in Table \ref{tab:bma}. In addition to variables selected in the preferred model, variables which have potential to explain injurious falls were also identified in other potential models. These were Tinetti gait, gender, and functional reach. In other models (not shown), profiles that were prone to have injurious falls are: being female, having relatively good balance and gait, good self-rated physical functioning and general health, being overweight or obese, being fearful to falls, and having a relatively higher level of anxiety. 

Using these models, the classification rates for injurious/non-injurious falls prediction are reasonably high. For the preffered model, 78\% accuracy, 81\% sensitivity and 76\% specificity were achieved. These rates are comparable to that of BMA results (Figure \ref{fig:roc2}).

\section {Discussion}
\label{sec:discussion}

We have demonstrated the classification of fall/non-fall, and injurious/non-injurious falls using a random effects logistic regression model. The models selected in the BMA approach showed that fearfulness, Tinetti gait, previous falls, Beck's anxiety and physical functioning of SF-36 were associated with fall/non-fall incidences. Although it is not selected in the model, a relatively high chance of falls for patients with relatively good self-rated balance and fearfulness (good, very good, and excellent for balance, not at all, slightly and moderately for fearful) obtained in the descriptive analysis raised some concerns on the reliability of patients self-rating assessment. This result might explain findings in \cite{Lamont2017}, that most falls occurred when medications were effectively working.

The reliability of the self-assessed health condition was also questioned in previous studies, for example, as in \cite{Crossley2002} since these researchers found that a considerably high proportion of respondents (28\%) changed their answer to their health status when asked twice. A more recent study also found a similar result, with a higher proportion of almost 40\% of respondents changing their answer about their general health condition when asked again one month later \citep{Zajacova2011}. 

There is also a similar concern for injurious falls, where patients with better balance, self-rated physical functioning and general health also have a higher risk of getting injured when they fall. It is recommended for the health carer of people with PD to elaborate on the outdoor activities of patients in the afternoon, when the high rates of injurious falls occurred. Attention is also needed to take steps on ``fall proofing'' inside the house, as the high rate of injurious falls also occurred in the house at night (around 10 p.m.). 

The SF-36 is designed for the coherent, self-rate health assessment of a general population, and therefore may not adequately capture aspects of the health related quality of life specific to PD. A measure of subjective health status specifically designed for PD is the PDQ-39 (Parkinson's Disease Questionnaire-39) and was shown to be internally reliable and valid \cite{Jenkinson1997}. Nevertheless, \cite{Brown2009} showed that the SF-36 is more responsive than the PDQ-39 in identifying the change in patients health condition, and thus recommended a combination of these two measures to assess the health-related quality of life of people with PD. 

The limitation of this study is that it has used the SF-36 instead of the PDQ-39, or a combination of the SF-36 and the PDQ-39, as recommended in \cite{Brown2009} for the health-related quality of life of PD patients. Nevertheless, four subscales of the SF-36 measuring physical functioning, general health, energy/fatigue, and pain showed  a noticeable association with injurious falls. However, considering the question on the reliability of a self-assessed health questionnaire, this result should be interpreted with caution. Assessment based on an instrument more specific to PD, for example the PDQ-39, might provide a better insight in understanding injurious falls incidence. 

As there were positive association between injurious falls and body mass index, balance and gait, attaining a normal body mass index, and improving balance and gait are recommended preventive efforts for injurious falls. There were no significant differences in the risk of falls between males and females, yet if falls occurred, females were more likely to get injured than males, confirming the findings in \cite{Wielinski2005}. 
 
\section{Summary}  
\label{sec:summary}

Through this study we identified risk factors for falls, and also for injurious falls. Fearfulness, Tinetti gait and previous falls were significant risk factors for falls. While fall time, location, fearfulness and functional reach were strongly associated with injurious falls.  Incorporating the result from BMA, attaining normal body mass index, improving balance and gait, reducing the level of anxiety and paying attention to taking precautions with outdoor activities during the day were recommended to reduce the risk of injurious falls. There were no significant differences in the risk of falls between males and females, yet if falls occurred, females were more likely to get injured than males. This finding may motivate research in health related areas to understand why females are more prone to injuries than males, and how to overcome this problem.
 
\bibliography{bibl_Factors_associated_with_injurious_falls_in_PD_Oct2019} 

\end{document}